\documentclass{physeauth}

\usepackage{graphicx}          
\usepackage{amsmath}           
\usepackage{amssymb}           

\begin{document}

\begin{frontmatter}
\title{Persistent current of correlated electrons in mesoscopic ring with impurity}

\author[IEE]{R. Kr\v{c}m\'{a}r},
\ead{roman.krcmar@savba.sk}
\author[IEE]{A. Gendiar},
\author[IEE]{M. Mo\v{s}ko},
\author[IEE]{R. N\'{e}meth},
\author[IEE]{P. Vagner},
\author[NCSU]{L. Mitas}

\address[IEE]{Institute of Electrical Engineering, Centre of Excellence CENG, Slovak Academy of
Sciences, 84104~Bratislava, Slovakia}
\address[NCSU]{Department of Physics, North Carolina State University,
Raleigh, NC 27695}

\begin{abstract}
The persistent current of correlated electrons in a continuous
one-dimensional ring with a single scatterer is calculated by
solving the many-body Schrodinger equation for several tens of
electrons interacting via the electron-electron (e-e) interaction
of finite range. The problem is solved by the
configuration-interaction (CI) and diffusion Monte Carlo (DMC)
methods. The CI and DMC results are in good agreement. In both
cases, the persistent current $I$ as a function of the ring length
$L$ exhibits the asymptotic dependence $I \propto L^{-1-\alpha}$
typical of the Luttinger liquid, where the power $\alpha$ depends
only on the e-e interaction. The numerical values of $\alpha$
agree with the known formula of the renormalisation-group theory.
\end{abstract}

\end{frontmatter}

%


A one-dimensional (1D) interacting electron system cannot be
viewed as a Fermi liquid of non-interacting quasi-particles. Away
from the charge-density-wave instability, the interacting 1D
system is a Luttinger liquid in which the elementary excitations
are of collective Bosonic nature (see the review \cite{Voit}). The
Luttinger liquid properties affect the 1D electron transport. An
interesting example of such effect is the persistent electron
current in a mesoscopic 1D ring. Magnetic flux $\phi$ piercing the
mesoscopic ring gives rise to the persistent current which can be
expressed (at $T = 0$K) as \cite{Imry-book}
\begin{equation} \label{Eqs-Vseob-Prud-a}
   I= -\partial E_0(\phi)/\partial \phi,
\end{equation}
where $E_0$ is the eigenenergy of the many-body groundstate. If the
ring is clean and the single-particle dispersion law is parabolic,
the electron-electron (e-e) interaction does not affect the
persistent current owing to the Galilean invariance of the
problem.
However, if a single scatterer (an impurity or a
weak link) is introduced into the ring, the non-interacting and
interacting result are expected to differ fundamentally. For
non-interacting spinless electrons in the 1D ring containing a
single scatterer with transmission probability
$|\tilde{t}_{k_{F}}|^2 \ll 1$, the resulting persistent current
depends on the magnetic flux and ring length ($L$) as
\cite{Gogolin}
\begin{equation} \label{I-nonint-approx}
I = (ev_F/2 L) \ |\tilde{t}_{k_{F}}| \ \sin(\phi'),
\end{equation}
where $\phi' \equiv 2\pi \phi/\phi_0$, $\phi_0 = h/e$ is the flux
quantum, $k_{F}$ is the Fermi wave vector, and $v_F$ is the Fermi
velocity. For a spinless Luttinger liquid the persistent current
follows the power law $I \propto L^{-\alpha-1}$. More precisely,
\cite{Gogolin}
\begin{equation} \label{I-Luttinger}
I \propto L^{-\alpha-1} \sin(\phi'),
\end{equation}
where  the power $\alpha$ depends only on the e-e interaction, not
on the properties of the scatterer. The formula
\eqref{I-Luttinger} can also be obtained \cite{Gogolin}
heuristically as follows. Matveev et al. \cite{Matveev} analyzed,
how the e-e interaction renormalizes the bare transmission
amplitude $\tilde{t}_{k_{F}}$ of a single scatterer in the middle
of the 1D wire. They derived the renormalized amplitude
$t_{k_{F}}$ by using the renormalization-group (RG) approach
suitable for a weakly-interacting electron gas ($\alpha \ll 1$).
In the limit $|\tilde{t}_{k_{F}}|^2 \ll 1$, their result can be
expressed in the form
\begin{equation} \label{transmission-Matveev}
t_{k_{F}} \simeq \tilde{t}_{k_{F}} (d/L)^{\alpha},
\end{equation}
where $L$ is the wire length, $d$ is the spatial range of the pair
e-e interaction $V(x-x')$, and the power $\alpha$ is given (for
spinless electrons) by expression
\begin{equation} \label{power-Matveev}
\alpha = [V(0)-V(2k_F)]/2\pi \hbar v_F,
\end{equation}
with $V(q)$ being the Fourier transform of $V(x-x')$. If one
replaces in equation \eqref{I-nonint-approx} the bare
$\tilde{t}_{k_{F}}$ by the renormalized amplitude
\eqref{transmission-Matveev}, one recovers the Luttinger-liquid
dependence \eqref{I-Luttinger}. This approach is heuristic and
single-particle, nevertheless, $\alpha$ is now expressed
microscopically in the form \eqref{power-Matveev}, expected to
hold for $\alpha \ll 1$.

 In the Luttinger-liquid model, the physics of the
low-energy excitations is mapped onto an effective field theory
using Bosonization \cite{Gogolin}, where terms expected to be
negligible at low energies are omitted. Within this model, the
asymptotic dependence \eqref{I-Luttinger} was obtained by another
approximation, using the analogy to the problem of quantum
coherence in dissipative environment \cite{Gogolin}. It would thus
be useful to avoid this approximation as well as Bosonisation.
This has been done in Ref. \cite{Meden}, where the persistent
current was calculated numerically by solving the 1D lattice model
with nearest-neighbor hopping and interaction. The formula
\eqref{I-Luttinger} was confirmed for long chains and strong
scatterers.

\begin{figure}[t]
\centerline{\includegraphics[clip,width=\columnwidth]{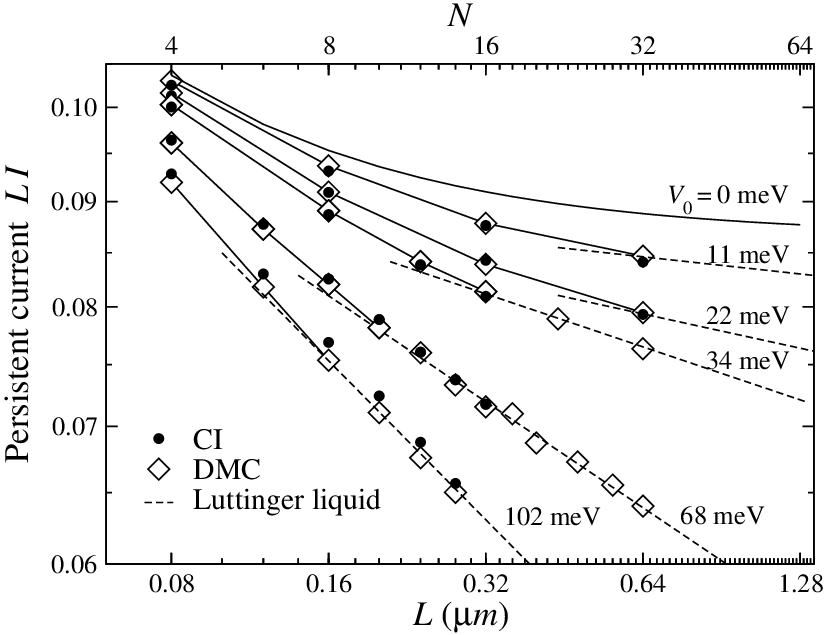}}
\vspace{-0.15cm} \caption{Persistent current $LI(\phi =
0.25\phi_0)$ in the GaAs ring with a single scatterer as a
function of the ring length $L$. The upper horizontal axis shows
the number of electrons $N$, the electron density is $N/L=5 \times
10^7$ m$^{-1}$. The transmission of the scatterer is
$|\tilde{t}_{k_{F}}|^2 = 0.03$. The range of the e-e interaction
is $d = 3$nm, the magnitude $V_0$ is varied. The CI and DMC data
are shown by symbols. The full lines connecting the DMC data are a
guide for eye. The dashed lines show the Luttinger liquid
asymptotics $LI \ \propto L^{-\alpha}$, with $\alpha$ obtained
from the RG formula \eqref{power-Matveev}: $\alpha=0.0277, 0.0561,
0.0855, 0.1710$, and $0.2565$ for $V_0 = 11, 22, 34, 68$, and
$102$~meV, respectively.} \label{Fig:2}
\end{figure}

\begin{figure}[t]
\centerline{\includegraphics[clip,width=\columnwidth]{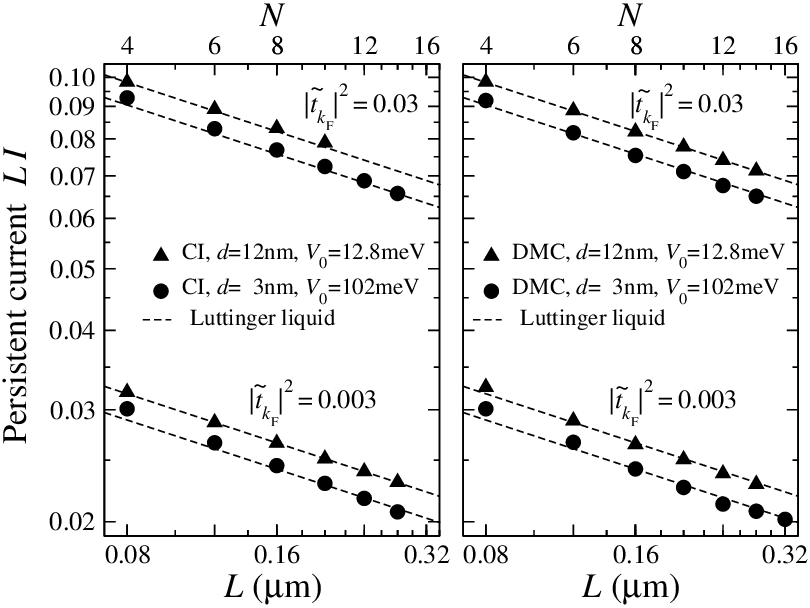}}
\vspace{-0.15cm} \caption{Persistent current $LI(\phi =
0.25\phi_0)$ in the GaAs ring with a single scatterer as a
function of the ring size. The parameters of the calculation are
given in the figure. The CI and DMC data are shown by symbols. The
dashed lines show the decay $LI \ \propto L^{-\alpha}$, with
$\alpha$ given by the RG formula \eqref{power-Matveev}: $\alpha =
0.2565$ for $d = 3$nm and $V_0 = 102$meV as well as  for $d =
12$nm and $V_0 = 12.8$meV. The CI and DMC data confirm this
$\alpha$. They also confirm that $\alpha$ remains unchanged when
the transmission $|\tilde{t}_{k_{F}}|^2$ of the scatterer is
varied.} \label{Fig:3}
\end{figure}

In this work the persistent current of correlated electrons is
studied in a continuous 1D ring, not in the 1D lattice. Using the
configuration-interaction (CI) and diffusion Monte Carlo (DMC)
methods, we solve the continuous many-body Schrodinger equation
for several tens of electrons interacting via the e-e interaction
\begin{equation} \label{VeeExp}
V(x - x') = V_0 \,  \exp(- \left| x - x' \right|/d).
\end{equation}
Interaction \eqref{VeeExp} emulates screening and allows
comparison with correlated models \cite{Gogolin,Matveev,Meden}
which also assume the e-e interaction of finite range. The figures
\ref{Fig:2} and \ref{Fig:3} present our major CI and DMC results.
These results show the persistent current exhibiting the
Luttinger-liquid law $I \propto L^{-1-\alpha}$, already verified
by the lattice model \cite{Meden}. However, it has sofar not been
verified, that the power $\alpha$ obeys the RG formula
\eqref{power-Matveev} also in the microscopic model avoiding the
RG approach. The CI and DMC data in figures \ref{Fig:2} and
\ref{Fig:3} verify this fact quite well.

We consider $N$ electrons with free motion along a circular 1D
ring threaded by magnetic flux $\phi$ due to the constant magnetic
field. The non-interacting single-electron states $\varphi_i(x)$
obey the Schr\"odinger equation
\begin{equation}\label{1-e}
 \left(-\frac{\hbar^2}{2m}\frac{d^2}{dx^2} + \gamma\delta(x)\right)\varphi_i(x) =
 \epsilon_i\varphi_i(x),
\end{equation}
with boundary condition $\varphi_i(x + L) =
\exp\left(i\frac{2\pi\phi}{\phi_0}\right)\varphi_i(x)$, where $x$
is the electron coordinate along the ring, $m$ is the electron
effective mass, and $\gamma\delta(x)$ is the potential barrier due
to the scatterer. The wave functions $\varphi_i(x)$ and energy
levels $\epsilon_i$ are found numerically as in Ref.
\cite{Nemeth}. We want to solve the Schr\"odinger equation $H \Psi
= E \Psi$ for $N$ interacting electrons with Hamiltonian
\begin{equation}\label{MBS}
H\! =\! \sum_{i=1}^N\left(\!-\frac{\hbar^2}{2m}\frac{d^2}{dx_i^2}
+ \gamma\delta(x_i)\!\!\right) + \frac12\sum_{i \neq j}V(x_i -
x_j).
\end{equation}
Consider first the non-interacting many-body problem
$\mathcal{H}\chi = \mathcal{E}\chi$, where $\mathcal{H}$ is
Hamiltonian \eqref{MBS} without the e-e interaction. Clearly, in
this case $\mathcal{E}_i=\epsilon_{i_1}+\dots+\epsilon_{i_N}$ and
\begin{equation}\label{SLater}
 \chi_i = \frac{1}{\sqrt{N}}
        \left|
        \begin{array}{ccc}
        \vspace{-0.89cm}
        \phantom{\vdots} & & \\
        \varphi_{i_1}(x_1) & \dots & \varphi_{i_N}(x_1)\\[-0.22cm]
        \vdots & \ddots & \vdots\\[-0.26cm]
        \varphi_{i_1}(x_N) & \dots & \varphi_{i_N}(x_N)
        \vspace{-0.08cm}
\end{array}\right|,
\end{equation}
where $i$ is the quantum number labeling a specific set of $N$
occupied levels $\epsilon_{i_1},\dots,\epsilon_{i_N}$. The CI method
\cite{Foulkes-01} relies on expansion
 $\Psi = c_0\chi_0 + c_1\chi_1 + c_2\chi_2 + \dots$. Using this expansion and equation $\langle\chi_i|H|\Psi\rangle =
\langle\chi_i|E|\Psi\rangle$ we get
\begin{equation} \label{infiniteset}
\sum_{j=0}^\infty\left(\mathcal{E}_j\delta_{ij} + V_{ij}\right)c_j
= Ec_i, \quad  i = 0,1, \dots, \infty,
\end{equation}
where $V_{ij} = \frac12\langle\chi_i|\sum V(x_k -
x_l)|\chi_j\rangle$. We reduce the infinite number of
single-energy levels $\epsilon_j$ to the finite one by introducing
a proper upper energy cutoff. This reduces the infinite number of
equations \eqref{infiniteset} to a certain finite number $M+1$. We
get the finite system $\sum_{j=0}^M\left(\mathcal{E}_j\delta_{ij}
+ V_{ij}\right)c_j = Ec_i$, where $i = 0, 1, \dots, M$. This
system determines the eigenvalues $E_l$ and eigenvectors
$(c^{l}_0, c^{l}_1, \dots, c^{l}_M)$ for $l = 0, 1, \dots, M$. To
obtain the persistent current \eqref{Eqs-Vseob-Prud-a}, we need
only the groundstate energy $E_{l=0}$ as a function of $\phi$. We
have obtained $E_0$ by solving the system with program ARPACK.
%
%

We have also calculated $E_0$ by the DMC method \cite{Foulkes-01}.
Technical details of this calculation are the same as in Ref.
\cite{Vagner}, the persistent current is obtained from
(\ref{Eqs-Vseob-Prud-a}).



\begin{figure}[t]
\centerline{\includegraphics[clip,width=\columnwidth]{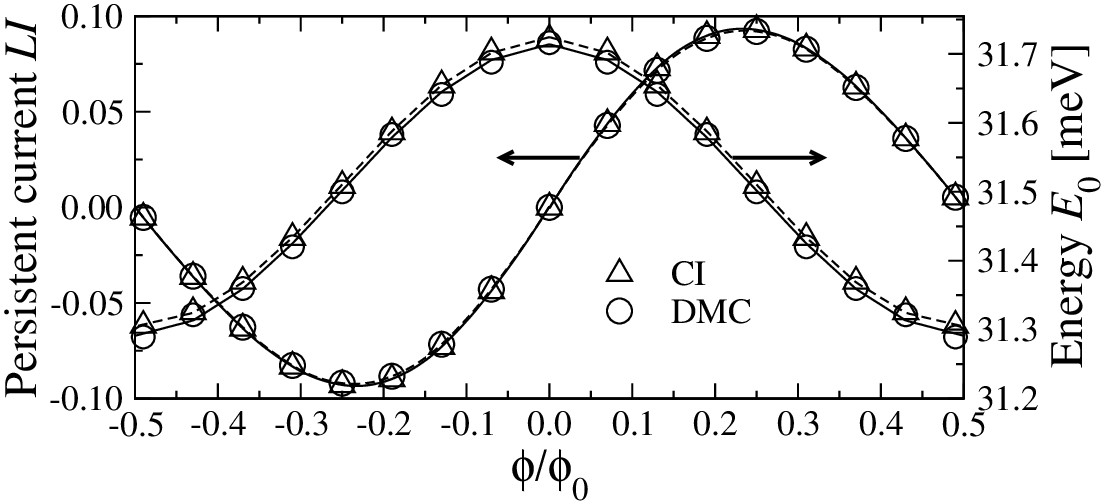}}
\vspace{-0.15cm} \caption{Persistent current versus magnetic flux
and energy of the many-body ground state for the GaAs ring with
parameters $N = 4$ and $L=0.08$ $\mu$m. The transmission of the
scatterer is set to $|\tilde{t}_{k_{F}}|^2 = 0.03$. The parameters
of the e-e interaction are $V_0 = 102$~meV and $d = 3$ nm. The CI
and DMC data are shown by open symbols, the lines connecting
points are a guide for eye. The data for the persistent current
follow the dependence $I(\phi) \propto \sin(2 \pi \phi/
\phi_0)$~\cite{Gogolin}.}
\label{Fig:1}
\end{figure}

Our major CI and DMC results are shown in figures \ref{Fig:2} and
\ref{Fig:3}. These figures show in log scale the persistent current
$LI(\phi = 0.25\phi_0)$ as a function of the ring length for a
GaAs ring with one strong scatterer. In log scale the decay $LI
\propto L^{-\alpha}$ is linear, with slope $-\alpha$. The CI and
DMC data in figure \ref{Fig:2} confirm the expected
\cite{Gogolin,Meden} trend: The stronger the e-e interaction the
shorter the system necessary to reach the $LI \propto L^{-\alpha}$
asymptotics. In figure \ref{Fig:3}, where only the strongest e-e
interaction is considered, the decay $LI \propto L^{-\alpha}$
exists already in rings with four electrons. To see this
Luttinger-liquid feature in such few-electron system is a
surprising result. Further, the CI and DMC data in figures
\ref{Fig:2} and \ref{Fig:3} confirm that $\alpha$ obeys the RG
formula \eqref{power-Matveev}. Finally, figure \ref{Fig:1} shows,
that the persistent current is a sine function of the magnetic
flux also in the CI and DMC models .


We thank for the grant APVV-51-003505, VEGA grant 2/6101/27 and ESF project VCITE at IEESAS.
L.M. thanks for support by NSF.


%
%

\end{document}